# On the local streamline pattern of planar polynomial velocity field with nonzero linear part

Jian Gao [1)], Hongping Ma [2)], Rong Wang [1)], Wennan Zou [*1),2)]

1) Institute of Fluid Mechanics, Nanchang University, Nanchang 330031, China
2) Institute for Advanced Study, Nanchang University, Nanchang 330031, China
*Authors to whom correspondence should be addressed: zouwn@ncu.edu.cn

**Abstract.** The streamline pattern of planar polynomial velocity field is far from fully understood. In the community of fluid mechanics, most studies simply focus on the velocity gradient, or the linear part of the velocity field, but few studies on high-order terms. This paper is concerned with the local streamline pattern (LSP) of velocity field around an isotropic point. In virtue of the concept and method of dynamical systems, where the streamline pattern is equivalent to the phase portrait, we make clear the classification of LSPs of planar velocity fields with nonzero linear part, especially the cases where the determinant of velocity gradient vanishes at the isotropic point.

## 1. Introduction

Our understanding of the flow of real fluids, such as air, water, etc., has greatly influenced our viewpoint about the world, while accurate prediction of complex flows, say turbulence, remains a challenge to human intelligence to this day (Frisch and Orszag, 1990 [1]; Barenblantt and Chorin, 1998 [2]). Velocity of fluid in motion is no doubt at all the statistical average of the velocities of a large number of fluid particulates centered at a point. The velocity field on the domain occupied by the fluid consists of a differentiable function in space and time. In order to support the understanding that the velocity can be used to describe the movement of identifiable pieces of matter, Batchelor (1970) [3] considered that the initial linear dimensions of fluid element must be so small as to guarantee smallness at all relevant subsequent instants in spite of distortions and extensions of the element. However, when we define pieces of fluid with any finite scale at a certain time in a flow, say the simplest Couette flow, the shear process will keep two smaller pieces of fluid, perpendicular to the flow direction, away from each other and, finally cannot be maintained at a limited distance. The aim of velocity analysis is to make clear the transport and contact in the fluid. From the viewpoint of community of fluid mechanics (Arnold and Khesin, 2001 [4]; Moffat, 2021 [5]), the topology aspects of fluid flow are all from the velocity, and not limited to streamline pattern. On the other hand, considering that fluid particulates cannot maintain their aggregation during the flow (Murdoch, 2012) [6], it is reasonable to regard the fluid flow as the mutual sliding of fluid layering at molecular level, and so the viscous friction as the response of the fluid slip strength, such that new fields in addition to velocity field could be introduced to describe the slip contact relationship of fluid forming and keeping active during the flow (Zou, 2016) [7]. Then, the flow field analysis will not be limited to velocity analysis, and the streamline pattern from velocity direction possesses more implication of slip structures in fluid.

Streamlines are lines in the fluid whose tangent is everywhere parallel to the instantaneous velocity; when the flow is steady, the streamlines have the same form at all times [3]. Streamline as an important concept to explain the resistance produced on a stationary body in a current of fluid, was alluded to by Daniel Bernoulli [8] in 1738 in the analysis for the impact of a jet against a plate (cf. Calero, 2008 [9]), and first described by Euler [10] in his commentaries to Robins' Gunnery, when he translated it into German in the year 1745 (cf. Truesdell, 1968 [11]). But the definitions of different scientists seem to be not exactly the same. Rankine (1871) [12] said that "a streamline is the line that is traced by a particle in a current of fluid. In a steady current each individual streamline preserves its figure and position unchanged, and marks the track of a filament or continuous series of particles that follow each other." In steady flows, streamlines are the same as trajectories and streak lines, and so can be used for different understandings. In this paper, the streamline is understood as the contact structures of fluid, even in unsteady (turbulent) flows, as a basic reference for the further analysis.

In the early 1950s, Robert Legendre introduced the basic concepts of the Critical Point Theory to provide a rational definition of separation in three-dimensional flows, while Henri Werlé carried out demonstrative experiments of stationary structures of separation zones (cf. Délery, 2001 [13]). Tobak and Peake (1982) [14] and Perry and Chong (1987) [15] reviewed the applications of streamline pattern based on critical point(s) and associated them with the theory of dynamical systems. These issue-oriented studies are almost of three-dimensional. Bakker (1991) [16] first systematically strengthened the applications of the qualitative theory of differential equations to flow patterns (Guckenheimer and Holmes, 1983 [17]), where only the planar system is fairly well understood. Thereafter, most developments (Brøns, 1994, 2007; Brøns and Hartnack, 1999; Brøns et al., 1999, 2001; and Hartnack, 1999a, 1999b; Jiménez-Lozano and Sen, 2010; Deliceoglu, 2013; Yokoyama and Sakajo, 2015; Sakajo and Yokoyama, 2018) [18-28] were concerned with 2D streamline topology, only a few were of in general three-dimensional (Bajer and Moffatt, 1990 [29]). As Brøns (2001) [30] pointed out, a structure is qualitatively rather than quantitatively defined, and some basic concepts in fluid mechanics are in fact naturally defined in dynamical systems terms.

Dynamics is a time-evolutionary process, while dynamical systems are generally described by differential or difference equations (Layek, 2015) [31]. Since analytical solution of nonlinear equations is difficult to obtain except in a few special cases, Poincaré (1886) [32] was the first in his study concerning the stability and evolution of the solar system to emphasize the qualitative approach to a system of differential equations

$$\frac{dx_1}{v_1} = \cdots = \frac{dx_n}{v_n} = dt, \qquad (1)$$

where $\{v_i, i = 1, \cdots, n\}$ are given functions, real and uniform depending on the state variables $\{x_i, i = 1, \cdots, n\}$, analytical with respect of these variables, and $t$ indicates the time. Birkhoff (1912) [33][34] originally presented the name of 'Dynamical Systems', while Andronov (Andronov et al., 1966) [35] contributed immensely to the mathematical theories for dynamical systems concerned with nonlinear oscillations and their applications. From then on, the advance of dynamical systems led to the development of a rich and powerful field with applications to physics, biology, meteorology, astronomy, economics and other areas (Brin and Stuck, 2001) [36], and made significant contributions to understanding some nonlinear phenomena. Most of the literature is limited to the polynomial autonomous (dynamical) system due to the following reasons:

- Polynomial is an excellent kind of analytical function, and many practical dynamical systems are of polynomial;
- For any dynamical system, the local properties of a critical point can be analyzed through the Taylor series expansion of the function $\{v_i, i = 1, \cdots, n\}$ at the point;
- As Lloyd (1988) [37] said, the simplicity of the polynomial statement of dynamical system belies the difficulty of making appreciable progress and, the study is so striking that the hypothesis is algebraic while the conclusion is topological;
- Any nonautonomous system can be transformed into an autonomous one by redefining time as a new dependent variable (Wiggins, 1990) [38], although it is not an autonomous system with critical points.

Three-dimensional (3D) polynomial autonomous systems are quite complicate, exhibiting all the types of behavior we have seen for two-dimensional systems, plus much more complicated behavior (Llibre and Teruel, 2014) [39]. Few authors talked about the classification of 3D linear autonomous systems (Arnold, 1992 [40]; Chong, et al., 1990 [41]; Verhulst, 1996 [42]; Hirsch et al., 2004 [43]; Zou et al., 2021 [44]). Lorenz (1963) [45] derived a three-dimensional quadratic system from a model of fluid convection in the atmosphere, and other researchers have found it to be relevant to a variety of phenomena from lasers to water wheels. Lorentz pointed out, convincingly but without rigorous proof, that for some values of coefficients the solutions of this system exhibit chaotic-like behavior, which is called the *strange attractor* (Ruelle and Takens, 1971) [46], having the fractional dimension between 2 and 3 (not being a point neither a curve nor a surface). The bifurcation study of the seminal Lorentz equation of quadratic systems may be a good example to illustrate the extreme complexity of polynomial systems (Kelley and Peterson, 2010) [47]. Bajer and Moffatt (1990) [29] studied a 3D quadratic Hamilton system whose streamlines are confined in a sphere. In fact, we do not have complete knowledge of behaviors of 3D nonlinear dynamical systems, even there is no consensus on how to describe their topological features. Relatively speaking, two-dimensional (2D) polynomial systems do not display the complicated dynamical features which can arise only in higher dimensions, and it is reasonable to expect to be able to describe their structures more complete (Lloyd, 1988) [37].

For the planar polynomial system

$$\dot{x}_1 = v_1(x_1, x_2), \qquad \dot{x}_2 = v_2(x_1, x_2), \tag{2}$$

in which $v_1$ and $v_2$ are polynomials of $x_1$ and $x_2$, and consist of a vector field as components, some important questions are still to be answered beyond the field of linear differential equations. For the system (1), the question to make clear the number of limit cycles (isolated periodic orbits inside the set of all periodic orbits) and their distribution in the plane, is the 16th of 23 problems posed by D. Hilbert at the Second International Congress of Mathematicians, Paris, in 1900, and is still unsolved (Ye, 1986 [48]; Li, 2003 [49]). Most notions of 2D dynamical systems can be generalized to higher dimensions (Dumortier *et al.*, 2006) [50]. The qualitative study of dynamical systems does not find an explicit expression of their solutions, but instead makes clear the phase portrait, including its description, classification (equivalence) and bifurcation when the equations change from one class to another (Llibre and Teruel, 2014) [39]. The streamline pattern (SP) is a variant name for the phase portrait when the vector field in (1) is recognized as the velocity field of fluid. In order to present the (full) phase portrait of a dynamical system, we must first find out all its (finite and infinite) critical points, their distribution, and the local phase portrait near every critical point.

This paper will focus on the local streamline pattern (LSP) of an isolated critical point of a 2D velocity field with nonzero linear part, including its classification and index calculation. In Section 2, we first introduce some concepts that are necessary to expand our analysis, such as isotropic point, index, separatrix, sectors, various transformations, etc., where the invariance under a series of transformations of spatiotemporal coordinates are used to define the qualitative classification, and in addition, the positive definite transformation of velocity is adopted to carry out the topological equivalence with the same index. In Section 3, the types of linear velocity field are revisited with the transformation definition, and in Section 4 we study the nonlinear velocity fields with nonzero linear part by detailed analyses of sectors, even by the polar blow-up technique. In Section 5 we develop a novel method to compute the index of velocity field at the isotropic point. Finally, some conclusions are listed.

## 2. Concept preparation

We divide the points in flow domain into the regular points with nonzero velocity and the singular points where the velocity vanishes. For a regular point, there is a unique streamline passing through it, while a singular point could be (1) isolated if no other singular points exist in one of its neighborhoods, (2) removable if the velocity directions of the regular points approaching it have a limit. In this paper, the polynomial expansions of velocity components around an isolated singular point must have no cofactor or be relatively prime. An *isotropic point* is defined as an isolated and unremovable singular point. It is well known that the LSP around an isotropic point could be very changeful and plays a key role in understanding the structure of velocity field (Legendre,1956 [51]; Tabak and Peake, 1982 [52]; Perry and Chong, 1987 [15]).

The streamline pattern in steady flows fully displays the flow structures, consisting of the LSPs and the distribution of the isotropic points. Although the transient streamline pattern of velocity field in unsteady flows (turbulence) does not fully correspond to the flow structures, such as the vortices entrained by the mainstream, small-scale vortices, etc., it is still the basis for understanding the mechanism of complex flows. Brøns (2007) [19] pointed out that a basic qualitative understanding of the streamline pattern in the flows is necessary, even if the final goal of treating a fluid mechanics problem is a quantitative one. And in the studies of differential equations and dynamical systems, a qualitative analysis is further used as a fundamental mean for general research. This paper will conduct a qualitative classification research on the LSP of an isotropic point in two-dimensional flows, with special emphasis on the complicated types determined by the nonlinear terms.

Assume that the velocity $(v_1, v_2)$ in (2) is analytic in the plane $(x_1, x_2)$ and takes the origin as one of its isotropic points, we can expand it as

$$v_1 = P_m(x_1, x_2) + o(r^m), v_2 = Q_n(x_1, x_2) + o(r^n), \tag{3}$$

where $r^2 = x_1^2 + x_2^2$, $P_m$, $Q_n$ are homogeneous polynomials of order $m, n \geq 1$. The equation (2) only including $P_m$ and $Q_n$ is called the principal equation at the isotropic point $(0,0)$ (Zhang, 1992) [53]. Two remarks on the qualitative analyses should be mentioned: (1) a cofactor that is larger than zero everywhere except at the origin doesn't affect the LSP; (2) higher order term(s) must be considered if $P_m$ and $Q_n$ cannot ensure the origin is an isotropic point. So, in general we assume $P_m$ and $Q_n$ to be relatively prime. The isotropic point is said to be linear or *elementary* if $m = n = 1$ and two eigenvalues of its coefficient matrix being nonzero; otherwise called nonlinear, or higher order. Among the nonlinear isotropic points, the cases without the linear term are called intricate, where the case with $m = n > 1$ is called homogeneous (Artés *et al.*, 2021) [54]. The isotropic point with term(s) more than the degree of homogeneous polynomials is called complex. Poincaré (cf. Zhang, 1992) [53] pointed out that a complex isotropic point can be considered as formed by coalescing several elementary isotropic points. Thus, due to a nonlinear isotropic point having the same qualitative structure as a linear isotropic point, the former is called after the latter, say node, focus, saddle, and center (Zhang, 1992) [53]. Two common misconceptions in streamline pattern analysis are (1) overemphasizing linear isotropic points over nonlinear ones, and (2) believing that linear degenerate isotropic points are trivial.

The index of an isotropic point is an excellent quantity for characterizing the topological equivalence of isotropic points with different velocity expansions, which can be indicated by an integer called the rotation number of velocity field. Let $\mathcal{L}$ be a piecewise smooth oriented close curve with the origin on its left, and passing through no isotropic points. Thus, $v_1^2 + v_2^2 \neq 0$ on $\mathcal{L}$, assume that the velocity makes an angle $\phi$ with the $x_1$-axis, the index $\text{Ind}(v, \mathcal{L})$ is defined by the rotation number (Grimshaw, 1990 [55]; Zhang, 1992 [53])

$$\text{Ind}(v, \mathcal{L}) = \frac{1}{2\pi} \oint_{\mathcal{L}} d\phi = \frac{1}{2\pi} \oint_{\mathcal{L}} \frac{v_1 dv_2 - v_2 dv_1}{v_1^2 + v_2^2}. \tag{4}$$

If the origin is the only isotropic point in the domain with the curve as its boundary, the index is also called the index of the isotropic point, denoted by $\text{Ind}_O$; Otherwise, if there are many isotropic points $s_1, s_2, \cdots$ in $\mathcal{L}$, we have $\text{Ind}(v, \mathcal{L}) = \sum_i \text{Ind}_{s_i}$. From (4), it is easy to prove that the index is independent of the speed distribution. If denote the index of an isotropic point by $J_O(v_1, v_2)$, Zhang (1992) [53] reported the following properties ($a \neq 0$):
$$J_O(x_1, x_2) = 1, J_O(v_2, v_1) = -J_O(v_1, v_2), J_O(av_1, v_2) = \text{sgn}(a) \cdot J_O(v_1, v_2), J_O(v_1 + v_2, v_2) = J_O(v_1, v_2). \tag{5}$$
We can merge the latter three properties into

$$J_O(a_{11}v_1 + a_{12}v_2, a_{21}v_1 + a_{22}v_2) = \text{sgn}(\det(A)) \cdot J_O(v_1, v_2), \tag{6}$$

where the transformation matrix $A = \begin{pmatrix} a_{11} & a_{12} \\ a_{21} & a_{22} \end{pmatrix}$ should be nondegenerate, namely $\det(A) \neq 0$.

There are more qualitative properties of the LSP other than the index. For a linear velocity field, the isotropic points, whatever center, focus or node, have the same index, but possess different features from the eigenvalue structure or real Schur form. We may distinguish: how the streamlines pass through the isotropic point? are the directions of these streamlines entering isotropic point the same (tangent to each other)? is there streamline tangent to the circle centered at the isotropic point? and so on. The LSP would become more delicate when the effect of nonlinear terms is considered. On the other hand, the qualitative properties can be identified from the types of curvilinear sectors, each of them divided by two characteristic directions (CDs, or half lines or rays) starting from the isotropic point, and their relative positions (Zhang, 1992) [53]. It should be pointed out that a streamline tending to the origin must tend to it spirally or along a CD (Zhang, 1992) [53]. Except for the type of a center (focus), there are at least two CDs tending to the isotropic point as $r \to 0$ ($t \to \infty$ or $t \to -\infty$) defined by

$$\frac{d\theta}{dt} = \frac{x_1 v_2 - x_2 v_1}{r^2} = 0, \tag{7}$$

or equivalently,

$$\tan \alpha = \frac{r d\theta}{dr} = \frac{x_1 v_2 - x_2 v_1}{x_1 v_1 + x_2 v_2} = \frac{I(r,\theta)}{M(r,\theta)} = \frac{G(\theta) + o(1)}{A(r,\theta)} = 0, \tag{8}$$

where direction angle $\theta$ comes from $z = x_1 + \iota x_2 = re^{\iota \theta}$ with $\iota = \sqrt{-1}$, $\alpha$ is the angle between the radius vector and the velocity at the point $P(r,\theta)$, as shown in Fig.1. If there exist $\delta > 0, K > 0$ such that $|A(r,\theta)| < K$ when $r < \delta$, it follows that $G(\theta_c) = 0$ is a necessary condition for $\theta = \theta_c$ being a characteristic line. Hence, let $P_m(x_1, x_2) = r^m X_m(\cos\theta, \sin\theta), Q_n(x_1, x_2) = r^n Y_n(\cos\theta, \sin\theta)$,

$$0 = G(\theta) = \begin{cases} \cos\theta Y_n(\cos\theta, \sin\theta) & , m > n \\ -\sin\theta X_m(\cos\theta, \sin\theta) & , m < n \\ \cos\theta Y_n(\cos\theta, \sin\theta) - \sin\theta X_m(\cos\theta, \sin\theta) & , m = n \end{cases} \quad (9)$$

is called the characteristic equation of (3) (Nemytskii and Stepanov, 1960 [56]; Zhang, 1992 [53]). In virtue of the property (6), the assumption $m \leq n$ is always available. For the isotropic point of an analytic vector field other than a center or a focus, when $\delta$ is small enough, there are no more than three kinds of sectors in the neighborhood bounded by $\ell = \{z: |z| = \delta\}$: elliptic, hyperbolic, and parabolic (Zhang, 1992 [53]; Cronin, 2007 [57]), as shown in Fig. 2. The numbers of different sectors satisfy the Bendixson formula

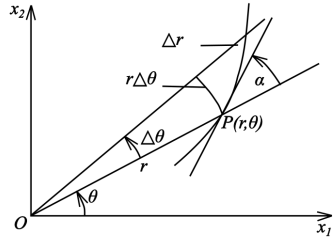

Figure 1. Included angle between the streamline and the position vector (cf. Zhang, 1992 [53]).

$$\text{Ind}(v, \ell) = 1 + \frac{e - h}{2} \quad (10)$$

where $e, h$ are the numbers of elliptic and hyperbolic sectors, respectively. The analysis of LSP can be carried out by the following steps: (1) find out all CDs (if there is no CD, the isotropic point is a center or a focus); (2) determine the type of sectors within all adjacent CDs; (3) judge the entry way of streamlines in a parabolic sector tending to the isotropic point. Usually, the separatrix is a radial half line defined by $\theta = \theta_c$, that is independent of the radial position, but sometimes when the lower order terms cannot define the isolated singular point, the separatrix may not be a radial line. Once the separatrices are confirmed, we can identify the flow directions on them by the signs of

$$\frac{d\ln r}{dt} = \frac{x_1 v_1 + x_2 v_2}{r^2}, \quad (11)$$

the streamline leaves out the singularity when it is greater than zero, or enters into the singularity when it is less than zero. Next, we can judge the sector to be parabolic if its two separatrices $2\pi \geq \theta_{c2} > \theta_{c1} \geq 0$ have the same streamline directions (Fig. 2(a)), and determine along which direction all streamlines in the sector tend in the isotropic point when $r \to 0^+$. If two separatrices of a sector have different streamline directions, say $\left.\frac{d\ln r}{dt}\right|_{\theta_{c1}} > 0 > \left.\frac{d\ln r}{dt}\right|_{\theta_{c2}}$, one can definitely find a tangent point of some streamline in the sector to the circle centered at the origin, namely $\left.\frac{d\ln r}{dt}\right|_{\theta=\theta_t} = 0$ with $\theta_t \in (\theta_{c1}, \theta_{c2})$, and $\left.\frac{d\theta}{dt}\right|_{\theta=\theta_t^+} < 0$ means a hyperbolic sector (Fig. 2(b)) while $\left.\frac{d\theta}{dt}\right|_{\theta=\theta_t^+} > 0$ means an elliptic sector (Fig. 2(c)). From the sectors and their types, we can calculate the index using he Bendixson formula (10).

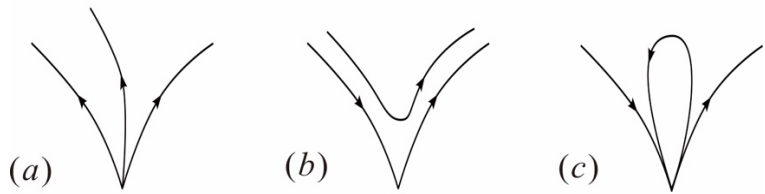

Figure 2. Three kinds of sectors in the neighborhood of an isotropic point: (*a*) parabolic, (*b*) hyperbolic, (*c*) elliptic.

The index of an isolated point can be used to build up a topological classification, and form an important base of streamline pattern, but it is not enough. Jiang and Llibre (2005) [58] proposed the qualitative equivalence of two isotropic points with the same index, that two streamlines start or end in the same direction at one isotropic point if and only if the equivalent two streamlines start or end in the same direction at the other isotropic point. For velocity fields with nonzero linear part, they presented well studied classification to date with some parameters, such as the eigenvalues of the coefficient matrix of linear part, and the leading term(s) of higher order. Artés *et al.* (2015) [59] proposed a finer classification called the geometric equivalence, with the algebraic tool of invariant polynomials. In the fluid mechanics community, most studies are limited to the nontrivial velocity gradient, where the coefficients of characteristic equation (Perry and Chong, 1987 [15]; Délery, 2001 [13]) and the real Schur form (Zou *et al.*, 2021) [44] are used to identify the streamline pattern.

Comparing with the qualitative equivalence and the geometric equivalence based on the algebraic parameters of velocity field, in this paper we try to introduce a novel classification of LSPs according to the invariance of velocity under the transformations of time, coordinates and velocity itself, together with the consideration of definite geometric features in the streamline pattern.

For an analytic velocity field with nonzero linear part in the neighborhood of an isotropic point, the following spatiotemporal transformations and linear transformation of velocity may be considered:

(i) the time transformation: $t \mapsto at, a \neq 0$, means: the absolute direction and speed of velocity don't affect the LSP.

(ii) the coordinate transformations,

(ii.1) the non-degenerate linear transformation: $\begin{pmatrix} x_1 \\ x_2 \end{pmatrix} \mapsto A \begin{pmatrix} x_1 \\ x_2 \end{pmatrix}, A = \begin{pmatrix} a_{11} & a_{12} \\ a_{21} & a_{22} \end{pmatrix}, \det A \neq 0$, which could be a rotation $\begin{pmatrix} \cos\phi & \sin\phi \\ -\sin\phi & \cos\phi \end{pmatrix}$, a deformation $\begin{pmatrix} \cos\varphi & \sin\varphi \\ -\sin\varphi & \cos\varphi \end{pmatrix}\begin{pmatrix} \alpha & 0 \\ 0 & \beta \end{pmatrix}\begin{pmatrix} \cos\varphi & -\sin\varphi \\ \sin\varphi & \cos\varphi \end{pmatrix}$ with $\alpha, \beta > 0$, a mirror $\begin{pmatrix} 1 & 0 \\ 0 & -1 \end{pmatrix}$, or their combination; that means the affine transformation doesn't change the LSP, and the chirality will not be used to distinguish the LSP, say the right spiral is equivalent to the left spiral.

(ii.2) a kind of near-identity transformation (Arnold, 1988 [60]; Kahn and Zarmi, 1998 [61]; Nayfeh, 2011 [62]): $\begin{pmatrix} x_1 \\ x_2 \end{pmatrix} \mapsto \begin{pmatrix} x_1 + h_1(x_1, x_2) \\ x_2 + h_2(x_1, x_2) \end{pmatrix}$, where $h_1(x_1, x_2), h_2(x_1, x_2)$ are undetermined higher polynomials used to eliminate the nonlinear terms in $v_1, v_2$, resulting in the *normal form* of velocity field; that means all nonlinear velocity fields with nonzero linear part are local equivalent if they have the same normal form at the isotropic point.

(ii.3) the blow-up transformation: for example $\begin{pmatrix} x_1 \\ x_2 \end{pmatrix} \mapsto \begin{pmatrix} x_1 \\ x_1 u \end{pmatrix}$, the idea behind the blow-up technique is to replace the plane by a surface and the singular point $p$ on the plane by a line or by a circle on the surface such that the singular point $p$ can be ideally split into a finite number of simpler singularities $p_i$ on the line or the circle (Artés et al., 2021) [54].

(ii.4) the positive power transformation: $\begin{pmatrix} x_1 \\ x_2 \end{pmatrix} \mapsto \begin{pmatrix} x_1^a \\ x_2^b \end{pmatrix}$ with $a, b > 0$, which will be specially applied to the hyperbolic sector, where the streamlines tending to the isotropic point are finite.

(iii) the proper linear transformation of velocity: $\begin{pmatrix} v_1 \\ v_2 \end{pmatrix} \mapsto B \begin{pmatrix} v_1 \\ v_2 \end{pmatrix}, B = \begin{pmatrix} b_{11} & b_{12} \\ b_{21} & b_{22} \end{pmatrix}$ with $\det B > 0$.

Besides these, it is well known that the streamline pattern is independent of the speed distribution. Applying these transformations to (2), we need no longer to distinguish attracting and repelling, left spiral and right spiral, and hyperbolic with and without expansion/compression. And for convenience, the nonzero expansion and compression parameters have no difference, the nonzero rotation parameter can be normalized to be one when no expansion/compression occurs. In general, the invariance under transformations (i) and (ii) yields a kind of qualitative classification, while the invariance under all transformations results in the topological classification.

## 3. Re-examination on the classification of linear velocity fields

Starting from the linear velocity field

$$\begin{pmatrix} \dot{x}_1 \\ \dot{x}_2 \end{pmatrix} = \begin{pmatrix} v_1 \\ v_2 \end{pmatrix} = D \begin{pmatrix} x_1 \\ x_2 \end{pmatrix}, D = \begin{pmatrix} d_{11} & d_{12} \\ d_{21} & d_{22} \end{pmatrix}, \tag{12}$$

we solve the characteristic equation $\begin{vmatrix} d_{11} - \lambda & d_{12} \\ d_{21} & d_{22} - \lambda \end{vmatrix} = 0$ to obtain two real eigenvalues $\lambda_1, \lambda_2$ or a couple of complex eigenvalues $\vartheta \pm \iota\mu$, and use the corresponding eigenvectors to construct an affine transformation $\begin{pmatrix} x_1 \\ x_2 \end{pmatrix} \mapsto A \begin{pmatrix} x_1 \\ x_2 \end{pmatrix}, \det(A) > 0$ so that

$$\begin{pmatrix} \dot{x}_1 \\ \dot{x}_2 \end{pmatrix} = J_D \begin{pmatrix} x_1 \\ x_2 \end{pmatrix} \tag{13}$$

with the Jordan matrix $J_D$ taking the form

$$J_D = \begin{pmatrix} \lambda_1 & 0 \\ 0 & \lambda_2 \end{pmatrix} \text{ or } \begin{pmatrix} \vartheta & \mu \\ -\mu & \vartheta \end{pmatrix} \text{ or } \begin{pmatrix} \vartheta & 1 \\ 0 & \vartheta \end{pmatrix}, \tag{14}$$

which can be applied to classify the LSP of linear velocity fields around the origin into six types (Jiang and Llibre, 2005) [58]: (1) saddle if $\lambda_1 \lambda_2 < 0$, (2) two-direction node if $\lambda_1 \lambda_2 > 0$, $\lambda_1 \neq \lambda_2$, (3) star-like node if $\lambda_1 = \lambda_2$ and $J_D$ can be diagonalized, (4) one-direction node if $\lambda_1 = \lambda_2 = \vartheta$ and $J_D$ cannot be diagonalized, (5) center if two eigenvalues are pure imaginary, (6) focus if two eigenvalues are complex with nonzero real part. For the cases of linear velocity fields with one or more zero eigenvalue(s), the origin is no longer an isolated singularity.

Recently, after a definite rotation, the real Schur form of $D$ ($\tau\omega \geq 0$)

$$S_D = RDR^T = \begin{pmatrix} \vartheta & \tau + \omega \\ \tau - \omega & \vartheta \end{pmatrix}, R \in SO(2) \tag{15}$$

is actively applied to work out some vortex criterions (Zhou et al., 1999 [63]; Li et al., 2014 [64]; Liu et al., 2018 [65]; Zou et al. 2021 [44]). Define $J_2 = \det(A) = \lambda_1 \lambda_2 = \vartheta^2 - \tau^2 + \omega^2$, we can also build up the classification by the parameter set $\{\vartheta, \tau, \omega\}$, namely (1) saddle if $J_2 < 0$, (2) two-direction node if $J_2 > 0$, $\tau^2 \neq \omega^2$, (3) starlike node if $\vartheta \neq 0$ and $\tau = \omega = 0$, (4) one-direction node if $\vartheta \neq 0$ and $\tau^2 = \omega^2 \neq 0$, (5) center if $\vartheta = 0, \tau^2 < \omega^2$, (6) focus if $\vartheta \neq 0, \tau^2 < \omega^2$.

The qualitative equivalence from the above two parameter systems is of algebraic nature (Artés et al., 2015) [59], but somehow incompletely conclusive. People may split one pattern into more through refined parameters or parameters of higher-order term(s). For example, sometime it is necessary to distinguish the expansion $\begin{pmatrix} 1 & 0 \\ 0 & 1 \end{pmatrix}$ and the compression $\begin{pmatrix} -1 & 0 \\ 0 & -1 \end{pmatrix}$, the clockwise vortex $\begin{pmatrix} 0 & -1 \\ 1 & 0 \end{pmatrix}$ and the counterclockwise one $\begin{pmatrix} 0 & 1 \\ -1 & 0 \end{pmatrix}$, the left spiral $\begin{pmatrix} 1 & -1 \\ 1 & 1 \end{pmatrix}$ and the right one $\begin{pmatrix} 1 & 1 \\ -1 & 1 \end{pmatrix}$; and it needs to explain why $\begin{pmatrix} 1 & 0 \\ 0 & -0.5 \end{pmatrix}$ and $\begin{pmatrix} 0 & 1 \\ 1 & 0 \end{pmatrix}$ belong to the same type. In this paper, based on the common intuition about the classification of local streamline pattern, we will adopt wider spatiotemporal coordinate transformations besides the usual scaling of time, space and rotation. Space rotation is used to carry out the real Schur form of the linear velocity field, while the deformation can be used to change the combination of $\tau$ and $\omega$ into a single $\tau$ if $\tau^2 > \omega^2$ or a single $\omega$ if $\tau^2 < \omega^2$ or $\tau + \omega = \pm 1$ if $\tau^2 = \omega^2$. Time inversion and scaling can be used to normalized $\vartheta = 1$, or $\omega = 1$ if $\vartheta = \tau = 0$, or $\tau = 1$ if $\vartheta = \omega = 0$. More transformations to be used for linear velocity fields include:

(i) the mirror transformation to unify $\begin{pmatrix} 1 & \omega \\ -\omega & 1 \end{pmatrix}$ and $\begin{pmatrix} 1 & -\omega \\ \omega & 1 \end{pmatrix}$, $\begin{pmatrix} 1 & 1 \\ 0 & 1 \end{pmatrix}$ and $\begin{pmatrix} 1 & -1 \\ 0 & 1 \end{pmatrix}$;

(ii) the positive power transformation to unify $\begin{pmatrix} \lambda_1 & 0 \\ 0 & \lambda_2 \end{pmatrix}$ with $\lambda_1 \lambda_2 < 0$ and $\begin{pmatrix} 1 & 0 \\ 0 & -1 \end{pmatrix}$.

Then, a complete standard form for linear velocity fields with $J_2 \neq 0$, called the quasi-real Schur form under the spatiotemporal coordinate transformations, can be built up, as listed in Table 1.

Finally, for the topological equivalence, by applying the positive definite linear transformation to the linear velocity fields, we can merge different qualitative types into two classes: $\begin{pmatrix} 1 & 0 \\ 0 & 1 \end{pmatrix}$ and $\begin{pmatrix} 1 & 0 \\ 0 & -1 \end{pmatrix}$, with the index values 1 and $-1$, respectively.

The classification of linear velocity fields can be achieved directly from their explicit solutions too. Substituting the quasi-real Schur form listed in Table 1 into (12), we derive the results as follows:

Case 1. saddle, from

$$\frac{d\ln r}{dt} = \frac{x_1 v_1 + x_2 v_2}{r^2} = \cos 2\theta, \qquad \frac{d\theta}{dt} = \frac{x_1 v_2 - x_2 v_1}{r^2} = -\sin 2\theta, \qquad (16)$$

we obtain the analytic solution satisfying $r(0) = r_0, \theta(0) = \theta_0$ to be

$$r(t) = r_0 \sqrt{e^{2t}\cos^2\theta_0 + e^{-2t}\sin^2\theta_0}, \qquad \theta(t) = \arctan(e^{-2t}\tan\theta_0). \qquad (17)$$

The qualitative properties follow from (17) include that: (1) the coordinate axes are streamlines and $\theta(t \to \pm\infty) \to \left(0^+, \frac{\pi^-}{2}\right)$ or $\left(\pi^+, \frac{3\pi^-}{2}\right)$ if $\tan\theta_0 > 0$, $\left(0^-, \frac{3\pi^+}{2}\right)$ or $\left(\pi^-, \frac{\pi^+}{2}\right)$ if $\tan\theta_0 < 0$, so there are four separatrices $\theta_c = 0, \frac{\pi}{2}, \frac{3\pi}{2}, \pi$, and no other streamline enters the isotropic point, say if $\theta_0 \neq \theta_c$, $r \to \infty$ when $t \to \pm\infty$; (2) for every circle centered at the origin, there are four points at $\theta_{1-4} = \frac{\pi}{4}, \frac{3\pi}{4}, \frac{5\pi}{4}, \frac{7\pi}{4}$, circumferential to four streamlines, respectively. The isotropic point with streamlines, except the separatrices, being hyperbolic is called a saddle, with the streamlines $x_1 x_2 = C$ and $\text{Ind}_O = -1$ from (12).

Table 1. Classification of elementary isotropic points

| Case | Determinant | Jordan form | Real Schur form | Quasi-real Schur form | Index | Sectors/Separatrices | Type |
|---|---|---|---|---|---|---|---|
| 1 | $J_2 < 0$ | $\lambda_1 > 0 > \lambda_2$ | $\tau^2 - \omega^2 > \vartheta^2 > 0$ | $\begin{pmatrix} 1 & 0 \\ 0 & -1 \end{pmatrix}$ | $-1$ | 4/4 | saddle |
| 2 | | $\lambda_1 > \lambda_2 > 0$ | $\vartheta^2 > \tau^2 - \omega^2 > 0$ | $\begin{pmatrix} 1+\tau & 0 \\ 0 & 1-\tau \end{pmatrix}, \tau \in (0,1)$ | 1 | 4/4 | two-direction node |
| 3 | | $\begin{pmatrix} 0 & \mu \\ -\mu & 0 \end{pmatrix}$ | $\vartheta = 0 > \tau^2 - \omega^2$ | $\begin{pmatrix} 0 & 1 \\ -1 & 0 \end{pmatrix}$ | 1 | 0/0 | vortex(center) |
| 4 | $J_2 > 0$ | $\begin{pmatrix} \vartheta & \mu \\ -\mu & \vartheta \end{pmatrix}$ | $\vartheta^2 > 0 > \tau^2 - \omega^2$ | $\begin{pmatrix} 1 & \omega \\ -\omega & 1 \end{pmatrix}, \omega > 0$ | 1 | 0/0 | spiral(focus) |
| 5 | | $\begin{pmatrix} \vartheta & 1 \\ 0 & \vartheta \end{pmatrix}$ | $\begin{pmatrix} \vartheta & 2\tau \\ 0 & \vartheta \end{pmatrix}$ | $\begin{pmatrix} 1 & 1 \\ 0 & 1 \end{pmatrix}$ | 1 | 2/2 | one-direction node |
| 6 | | $\begin{pmatrix} \vartheta & 0 \\ 0 & \vartheta \end{pmatrix}$ | $\begin{pmatrix} \vartheta & 0 \\ 0 & \vartheta \end{pmatrix}$ | $\begin{pmatrix} 1 & 0 \\ 0 & 1 \end{pmatrix}$ | 1 | $\infty/\infty$ | starlike node |
| 7 | $J_2 = 0$ | $\begin{pmatrix} \lambda & 0 \\ 0 & 0 \end{pmatrix}, \begin{pmatrix} 0 & 1 \\ 0 & 0 \end{pmatrix}$ | $\begin{pmatrix} \lambda & 0 \\ 0 & 0 \end{pmatrix}, \begin{pmatrix} 0 & 2\tau \\ 0 & 0 \end{pmatrix}$ | $\begin{pmatrix} 1 & 0 \\ 0 & 0 \end{pmatrix}, \begin{pmatrix} 0 & 1 \\ 0 & 0 \end{pmatrix}$ | - | - | straight |

Case 2. two-direction node, from

$$\frac{d\ln r}{dt} = (1+\tau)\cos^2\theta + (1-\tau)\sin^2\theta, \qquad \frac{d\theta}{dt} = -\tau\sin 2\theta, \tau \in (0,1) \qquad (18)$$

we have

$$r(t) = r_0 e^t \sqrt{e^{2\tau t}\cos^2\theta_0 + e^{-2\tau t}\sin^2\theta_0}, \qquad \theta(t) = \arctan(e^{-2\tau t}\tan\theta_0) \qquad (19)$$

with the following qualitative properties: (1) the coordinate axes are streamlines, $\theta(t \to \pm\infty) \to \left(0^+, \frac{\pi^-}{2}\right)$ or $\left(\pi^+, \frac{3\pi^-}{2}\right)$ if $\tan\theta_0 > 0$, $\left(0^-, \frac{3\pi^+}{2}\right)$ or $\left(\pi^-, \frac{\pi^+}{2}\right)$ if $\tan\theta_0 < 0$ and so there are four separatrices; (2) no streamline is tangent to the circle centered at the origin, the streamlines other than the separatrices enter the origin ($t \to -\infty$), and are tangent to radial lines with $\theta$ equal to $\frac{\pi}{2}$ or $\frac{3\pi}{2}$. The isotropic point with streamlines, except the separatrices, being parabolic is called a node, with the streamlines $x_2 = C x_1^{\frac{1+\tau}{1-\tau}}$ and $\text{Ind}_O = 1$ from (12).

Case 3. vortex, the explicit solution is simply $r(t) = r_0$ for streamline passing through $(r_0, \theta_0)$, all streamlines are circles except the center point, with the streamlines $x_1^2 + x_2^2 = C > 0$ and $\text{Ind}_O = 1$ from the definition.

Case 4. spiral, from

$$\frac{d\ln r}{dt} = 1, \qquad \frac{d\theta}{dt} = -\omega, \qquad (20)$$

we get $r(t) = r_0 e^t, \theta(t) = \theta_0 - \omega t$ for streamline passing through $(r_0, \theta_0)$, which means that there is no separatrix and all streamlines are helical starting from the origin, with the streamlines $\ln(x_1^2 + x_2^2)^\omega + \arctan\left(\frac{x_2}{x_1}\right) = C$, and $\text{Ind}_O = 1$.

Case 5. one-direction node, from

$$\frac{d\ln r}{dt} = (\cos^2\theta + \sin\theta\cos\theta) + \sin^2\theta = r(1 + \sin\theta\cos\theta), \qquad \frac{d\theta}{dt} = -\sin^2\theta, \tag{21}$$

we have

$$r(t) = r_0\sqrt{(\cos\theta_0 + t\sin\theta_0)^2 + \sin^2\theta_0}\, e^t, \qquad \theta(t) = \arctan\left(\frac{\tan\theta_0}{1 + t\tan\theta_0}\right). \tag{22}$$

The streamline pattern has properties: (1) only the $x_1$-axis forms two separatrices at $\theta(t \to -\infty) \to 0$ or $\pi$; (2) no streamline is tangent to the circle centered at the origin, the streamlines other than the separatrices approach the origin ($t \to -\infty$), and are tangent to radial lines with $\theta$ equal to $0^-$ or $\pi^-$; (3) the velocity component $v_1$ changes its sign at the point on the line $x_1 + x_2 = 0$. This is specially called one-direction node, with the streamlines $x_1 = x_2(\ln|x_2| + C)$, $\text{Ind}_O = 1$ from (12).

Case 6. starlike node, the explicit solution is simply $\theta(t) = \theta_0$, all streamlines $x_2 = Cx_1$ start from the origin ($t \to -\infty$).

Case 7. undefined, the singular point is not isolated, and the streamlines $x_2 = C$ are straight.

We summarize the analysis process as follows: (1) solve out the separatrices, that means $\frac{d\theta}{dt} = 0$ but $\frac{dr}{dt} \neq 0$ when $r \neq 0$; (2) find out whether the directions of flow in two adjacent separatrices are the same or not (both expansion/compression or not), for the former, the sector is parabolic, and (3) for the later, find out the tangency of streamline with a circle centered at the isotropic point, if the tangency is external, the sector is hyperbolic, else elliptic. Besides the above situations, the cases without separatrix are center-like, namely the isotropic point is a center if the streamline is circular, else a spiral; for the node cases, the number of separatrices and the entry way of streamlines approaching to the isotropic point can be used to build up the classification.

## 4. Classification of nonlinear velocity fields with nonzero linear part

Velocity field is in general nonlinear. Assume that under the transformation $\begin{pmatrix} x_1 \\ x_2 \end{pmatrix} \mapsto A \begin{pmatrix} x_1 \\ x_2 \end{pmatrix}$, the equation (2) is changed to the coordinates corresponding to eigenvalues $\lambda_1, \lambda_2$, which could be complex here, then we have

$$\begin{pmatrix} \dot{x}_1 \\ \dot{x}_2 \end{pmatrix} = \begin{pmatrix} \lambda_1 & 0 \\ 0 & \lambda_2 \end{pmatrix}\begin{pmatrix} x_1 \\ x_2 \end{pmatrix} + \begin{pmatrix} c_1^{N,0} & \cdots & c_1^{m_1,m_2} & \cdots & c_1^{0,N} \\ c_2^{N,0} & \cdots & c_2^{m_1,m_2} & \cdots & c_2^{0,N} \end{pmatrix} \begin{pmatrix} x_1^N \\ \vdots \\ x_1^{m_1} x_2^{m_2} \\ \vdots \\ x_2^N \end{pmatrix} + o(r^N), N > 1. \tag{23}$$

According to the Poincaré-Dulac theorem, we can introduce a near-identity transformation (Kahn and Zarmi, 1998) [61] or a formal diffeomorphism (Arnold, 1988) [60]

$$\begin{pmatrix} x_1 \\ x_2 \end{pmatrix} \mapsto \begin{pmatrix} x_1 \\ x_2 \end{pmatrix} + \begin{pmatrix} s_1^{N,0} & \cdots & s_1^{m_1,m_2} & \cdots & s_1^{0,N} \\ s_2^{N,0} & \cdots & s_2^{m_1,m_2} & \cdots & s_2^{0,N} \end{pmatrix} \begin{pmatrix} x_1^N \\ \vdots \\ x_1^{m_1} x_2^{m_2} \\ \vdots \\ x_2^N \end{pmatrix} \tag{24}$$

to eliminate the nonlinear terms $c_i^{m_1,m_2}, i = 1,2$ by setting

$$s_i^{m_1,m_2} = \frac{c_i^{m_1,m_2}}{m_1\lambda_1 + m_2\lambda_2 - \lambda_i} \tag{25}$$

if $m_1\lambda_1 + m_2\lambda_2 - \lambda_i \neq 0$; this process can continue for next higher order part, and so on. The terms $x_1^{m_1} x_2^{m_2}$ with indices satisfying $m_1\lambda_1 + m_2\lambda_2 = \lambda_i$ is called to be resonant with the linear part, and must be retained for inspection. The nonlinear equation (23) without non-resonant terms is called the normal form of the velocity with such a type of linear part (Arnold, 1988 [60]; Nayfeh, 2011[62]). Besides the above general derivation, we detail the cases with zero eigenvalues as follows.

The case with one eigenvalue being zero is called semi-elementary. Assume that all terms with powers less than $N$ have been simplified to be of normal form, we begin to simplify the terms of order $N$ from the expression

$$\begin{pmatrix} \dot{x}_1 \\ \dot{x}_2 \end{pmatrix} = \begin{pmatrix} 1 & 0 \\ 0 & 0 \end{pmatrix}\begin{pmatrix} x_1 \\ x_2 \end{pmatrix} + \sum_{1<k<N} \begin{pmatrix} g_1^k(x_1,x_2) \\ g_2^k(x_1,x_2) \end{pmatrix} + \begin{pmatrix} p_1^N(x_1,x_2) \\ p_2^N(x_1,x_2) \end{pmatrix} + o(r^N). \tag{26}$$

Making use of the transformation

$$\begin{pmatrix} x_1 \\ x_2 \end{pmatrix} \mapsto \begin{pmatrix} x_1 \\ x_2 \end{pmatrix} + \begin{pmatrix} h_1^N(x_1,x_2) \\ h_2^N(x_1,x_2) \end{pmatrix}, \tag{27}$$

Substituting it into (26) yields

$$\left[\begin{pmatrix} 1 & 0 \\ 0 & 1 \end{pmatrix} + \begin{pmatrix} \partial_1 h_1^N & \partial_2 h_1^N \\ \partial_1 h_2^N & \partial_2 h_2^N \end{pmatrix}\right]\begin{pmatrix} \dot{x}_1 \\ \dot{x}_2 \end{pmatrix} = \begin{pmatrix} 1 & 0 \\ 0 & 0 \end{pmatrix}\begin{pmatrix} x_1 + h_1^N \\ x_2 + h_2^N \end{pmatrix} + \sum_{1<k<n} \begin{pmatrix} g_1^k \\ g_2^k \end{pmatrix} + \begin{pmatrix} p_1^N \\ p_2^N \end{pmatrix} + o(r^N). \tag{28}$$

Due to

$$\left[\begin{pmatrix} 1 & 0 \\ 0 & 1 \end{pmatrix} + \begin{pmatrix} \partial_1 h_1^N & \partial_2 h_1^N \\ \partial_1 h_2^N & \partial_2 h_2^N \end{pmatrix}\right]^{-1} = \begin{pmatrix} 1 & 0 \\ 0 & 1 \end{pmatrix} - \begin{pmatrix} \partial_1 h_1^N & \partial_2 h_1^N \\ \partial_1 h_2^N & \partial_2 h_2^N \end{pmatrix} + o(r^{2N-2}) \tag{29}$$

and $2N - 2 \geq N$, we have

$$\begin{pmatrix} \dot{x}_1 \\ \dot{x}_2 \end{pmatrix} = \begin{pmatrix} 1 & 0 \\ 0 & 0 \end{pmatrix}\begin{pmatrix} x_1 \\ x_2 \end{pmatrix} - \begin{pmatrix} \partial_1 h_1^N & \partial_2 h_1^N \\ \partial_1 h_2^N & \partial_2 h_2^N \end{pmatrix}\begin{pmatrix} 1 & 0 \\ 0 & 0 \end{pmatrix}\begin{pmatrix} x_1 \\ x_2 \end{pmatrix} + \sum_{1<k<N}\begin{pmatrix} g_1^k \\ g_2^k \end{pmatrix} + \begin{pmatrix} p_1^N + h_1^N \\ p_2^N \end{pmatrix} + o(r^N), \tag{30}$$

and so, the simplification of order $N$ does not affect the terms with orders less than $N$, and vice-versa. The route of simplification is

$$p_1^N + h_1^N - x_1 \partial_1 h_1^N \to g_1^N, \qquad p_2^N - x_1 \partial_1 h_2^N \to g_2^N, \tag{31}$$

hence, the terms in $\begin{pmatrix} p_1^N \\ p_2^N \end{pmatrix}$ that cannot be eliminated from any $N$th-order homogeneous polynomials $\begin{pmatrix} h_1^N \\ h_2^N \end{pmatrix}$ are

$$g_1^N = a_{1,N-1}^N x_1 x_2^{N-1}, g_2^N = a_{2,N}^N x_2^N, \tag{32}$$

and finally, the normal form of (26) should be

$$\begin{pmatrix} \dot{x}_1 \\ \dot{x}_2 \end{pmatrix} = \begin{pmatrix} x_1 + x_1 \phi_1(x_2) \\ \phi_2(x_2) \end{pmatrix}, \tag{33}$$

where $\phi_1(x)$ is an arbitrary polynomial of orders greater than 0, $\phi_2(x)$ is an arbitrary polynomial of orders greater than one.

Table 2. Normal forms of velocity fields with nonzero linear part

| Determinant | Quasi-real Schur form | Index | Type | Resonant terms |
|---|---|---|---|---|
| $J_2 < 0$ | $\begin{pmatrix} 1 & 0 \\ 0 & -1 \end{pmatrix}$ | $-1$ | saddle | $\begin{pmatrix} x_1 \phi(x_1 x_2) \\ x_2 \phi_1(x_1 x_2) \end{pmatrix}$ |
| $J_2 > 0$ | $\begin{pmatrix} 1+\tau & 0 \\ 0 & 1-\tau \end{pmatrix}, \tau = \frac{n-1}{n+1}$ | 1 | two-direction node | $-$ or $\begin{pmatrix} ax_2^n \\ 0 \end{pmatrix}$ |
| | $\begin{pmatrix} 0 & 1 \\ -1 & 0 \end{pmatrix}$ | 1 | vortex or spiral | $\phi(r^2)\begin{pmatrix} x_1 \\ x_2 \end{pmatrix}$ |
| | $\begin{pmatrix} 1 & \omega \\ -\omega & 1 \end{pmatrix}, \omega > 0$ | 1 | spiral | - |
| | $\begin{pmatrix} 1 & 1 \\ 0 & 1 \end{pmatrix}$ | 1 | one-direction node | - |
| | $\begin{pmatrix} 1 & 0 \\ 0 & 1 \end{pmatrix}$ | 1 | star-like node | - |
| $J_2 = 0$ | $\begin{pmatrix} 1 & 0 \\ 0 & 0 \end{pmatrix}$ | - | ? | $\begin{pmatrix} x_1 \phi_1(x_2) \\ \phi_2(x_2) \end{pmatrix}$ |
| | $\begin{pmatrix} 0 & 1 \\ 0 & 0 \end{pmatrix}$ | - | ? | $\begin{pmatrix} 0 \\ \phi_2(x_1) + x_2 \phi_1(x_1) \end{pmatrix}$ |

Note: $a$ are real constant, $\phi(x)$, $\phi_1(x)$ indicate any two polynomials without constant term, while $\phi_2(x)$ is a polynomial of orders no less than 2.

The case with two eigenvalues being zero is called nilpotent. Similar derivation yields simplification relations

$$p_1^N + h_2^N - x_2 \partial_1 h_1^N \to g_1^N, \qquad p_2^N - x_2 \partial_1 h_2^N \to g_2^N \tag{34}$$

which result in normal form

$$\frac{d}{dt}\begin{pmatrix}x_1\\x_2\end{pmatrix}=\begin{pmatrix}x_2\\\phi_2(x_1)+x_2\phi_1(x_1)\end{pmatrix}. \tag{35}$$

It should be noticed that the normal form (35) is not unique, $\begin{pmatrix}x_2+x_1\phi_1(x_1)\\\phi_2(x_1)\end{pmatrix}$ is another choice (Wiggins, 1990, 2003) [38].

The normal forms of all nonlinear velocity fields with nonzero linear part are listed in Table 2. For the saddle type, the amplitudes of two eigenvalues of different signs could be adjusted by the positive power transformation, but do not affect the streamline pattern, Zhang (1992) [53] also proved that the additional resonant terms of higher order will not change the topological or qualitative structures of velocity fields near the isotropic point when their linear parts have eigenvalues with nonzero real part, i.e., the isotropic point is a saddle, a focus, or any kind of node; for the center case, where the eigenvalues being pure imaginary, the nonzero real part of $\phi(r^2)$ may change a center to a focus. Thus, the complexity mainly comes from the cases of semi-elementary and nilpotent with $J_2 = 0$, which will be discussed in the following.

For the semi-elementary case with a zero eigenvalue, since the positive cofactor of velocity components can be merged with time, the normal form equation

$$\dot{x}_1 = x_1[1+\phi_1(x_2)], \dot{x}_2 = \phi_2(x_2) \tag{36}$$

is qualitatively equivalent to

$$\dot{x}_1 = x_1, \dot{x}_2 = \frac{\phi_2(x_2)}{1+\phi_1(x_2)} = \phi_2'(x_2) = ax_2^m + o(x_2^m) \sim ax_2^m, a \neq 0 \tag{37}$$

in a small neighborhood of the origin, where $1+\phi_1(x_2)$ is always larger than zero. Making use of a scaling transformation (plus a rotation by 180 degrees if $c<0$)

$$x_1 \to cx_1, x_2 \to cx_2, c = \begin{cases} a^{1-m} &, m \text{ is even,}\\ |a|^{1-m} &, m \text{ is odd,}\end{cases} \tag{38}$$

we can simplify (36) to

$$\dot{x}_1 = x_1, \dot{x}_2 = \delta x_2^m, \delta = \begin{cases} 1 &, m \text{ is even,}\\ \pm 1 &, m \text{ is odd.}\end{cases} \tag{39}$$

Then, the local streamline pattern of an isotropic point can be determined by the sign of $a$ and the odevity of $m$ (Zhang, 1992) [53], as shown in Fig. 3. It becomes difficult to solve out the equations

$$\frac{d\ln r}{dt} = x_1v_1 + x_2v_2 = \cos^2\theta + \delta r^{m-1}\sin^{m+1}\theta, \frac{d\theta}{dt} = \frac{x_1v_2 - x_2v_1}{r^2} = (\delta r^{m-1}\sin^{m-1}\theta - 1)\sin\theta\cos\theta, \tag{40}$$

but for $r$ small enough, we know that $\frac{d\theta}{dt}=0$ will give four separatrices $\theta_{1-4} = 0, \frac{\pi}{2}, \pi, \frac{3\pi}{2}$, and so there are four sectors $s_{1-4}$ consecutively divided by the separatrices $\theta_1 - \theta_2 - \theta_3 - \theta_4 - \theta_1$. The flow directions of streamlines at $\theta_{2,4} = \frac{\pi}{2}, \frac{3\pi}{2}$ are determined by $\delta$ and $m$: (1) for $m$ being even, the separatrix $x_1=0, x_2>0$ is expansive, $s_1$ and $s_2$ are parabolic with starting streamlines tangent to the separatrix $x_1=0, x_2>0$, the isotropic point is called a saddle-node (Fig. 3(a)); (2) for $m$ being odd, all four sectors are parabolic with streamlines tangent to $x_2$-axis at the origin when $\delta=1$, or hyperbolic when $\delta=-1$, the isotropic point is called a node (Fig. 3(b)) or a saddle (Fig. 3(c)).

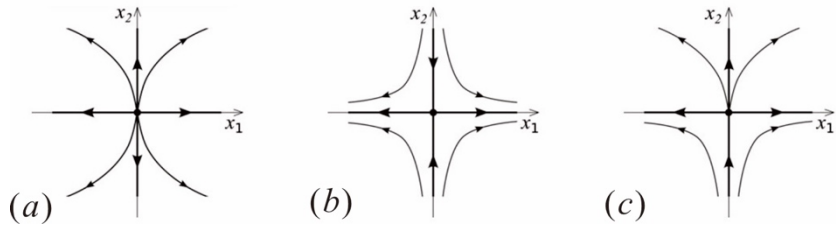

Figure 3. LSP of semi-elementary isotropic point: (*a*) saddle-node, (*b*) node, (*c*) saddle.

From the explicit solution $x_1 = x_{10}e^{\frac{1}{(m-1)\delta}\left(\frac{1}{x_{20}^{m-1}} - \frac{1}{x_2^{m-1}}\right)}$ of streamline passing through $(x_{10}, x_{20})$, we have

$$\frac{dx_1}{dx_2} = x_{10} e^{\frac{1}{(m-1)\delta x_{20}^{m-1}}} \left[ \delta x_2^m e^{\frac{1}{(m-1)\delta x_2^{m-1}}} \right]^{-1}. \tag{41}$$

for $\delta = 1$. Since

$$\lim_{x_2 \to 0} x_2^m e^{\frac{1}{(m-1)x_2^{m-1}}} \to +\infty \cdot \text{sign}(x_2) \text{ for } m = odd \text{ and } \lim_{x_2 \to 0^+} x_2^m e^{\frac{1}{(m-1)x_2^{m-1}}} \to +\infty \text{ for } m = even,$$

it is easy to prove that the streamlines are tangent to $x_2$-axis when the sectors are parabolic ($\delta = 1$). An interesting property of the hyperbolic sectors is the asymptotic behavior of streamlines as $x_2 \to \pm\infty$: it is obvious that all streamlines have asymptotic slope $\frac{dx_1}{dx_2}\big|_{x_2 \to \pm\infty} \to 0$, but $x_1 \to x_{10} e^{\frac{1}{(m-1)\delta x_{20}^{m-1}}} \neq 0$.

For the nilpotent case, the equation (2) becomes
$$\dot{x}_1 = x_2, \dot{x}_2 = \phi_2(x_1) + x_2\phi_1(x_1) = ax_1^p + bx_1^q x_2 + o(x_1^p + x_1^q x_2), a \neq 0, p > 1, q > 0. \tag{42}$$

Similar transformation to (38) can simplify (42) to
$$\dot{x}_1 = x_2, \dot{x}_2 = \delta x_1^p + bx_1^q x_2, \delta = \begin{cases} 1 & , p \text{ is even,} \\ \pm 1 & , p \text{ is odd.} \end{cases} \tag{43}$$

For (43), making use of the transformations $\begin{pmatrix} x_1 \\ x_2 \end{pmatrix} \mapsto \begin{pmatrix} 1 & 0 \\ 0 & -1 \end{pmatrix} \begin{pmatrix} x_1 \\ x_2 \end{pmatrix}, t \to -t$, we know that the sign of $b$ does not affect the LSP. Furthermore, the parameter value $b = 1$ can always be achieved by the transformations $\begin{pmatrix} x_1 \\ x_2 \end{pmatrix} \mapsto \begin{pmatrix} b^{\frac{2}{p-(2q+1)}} & 0 \\ 0 & b^{\frac{p+1}{p-(2q+1)}} \end{pmatrix} \begin{pmatrix} x_1 \\ x_2 \end{pmatrix}, t \to b^{\frac{1-p}{p-(2q+1)}} t$ if $p \neq 2q + 1$ and $b \neq 0$. Without loss of generality, we will take $b > 0$.

From
$$\frac{d\theta}{dt} = \frac{x_1 v_2 - x_2 v_1}{r^2} = (\delta r^{p-1} \cos^{p+1}\theta + br^q \sin\theta \cos^{q+1}\theta - \sin^2\theta), \tag{44}$$

we cannot determine any separatrix directly from the velocity field. Zhang (1992) studied this class of equations (43) by the blow-up technique, which originates from Poincaré, to decompose the nilpotent isotropic point into several elementary isotropic points, and concluded that the properties of the isotropic point will be determined by the size between $p$ and $2q + 1$, the odevity of $p$ and $q$, the sign of $\delta$ and the sign of $\Delta = b^2 + 4\delta(q + 1)$ for $p = 2q + 1$. Using the polar blow-up

$$x_1 = r\cos\theta \mapsto r^{\alpha+1}\cos\theta = r^\alpha x_1, x_2 = r\sin\theta \mapsto r^{\beta+1}\sin\theta = r^\beta x_2, dt = r^{\beta-\alpha-2}(r^2 + \alpha x_1^2 + \beta x_2^2)d\tau, \tag{45}$$

where the minimal values of powers $\alpha$ and $\beta$ will be introduced to consider the higher order effect, we obtain the following equations

$$\frac{dx_1}{dt} = \frac{r^{\beta-\alpha}(r^2 + \beta x_2^2) - \alpha\delta r^{\alpha p-\beta} x_1^{p+1} - \alpha b r^{\alpha q} x_1^{q+1} x_2}{r^2 + \alpha x_1^2 + \beta x_2^2} x_2, \tag{46.1}$$

$$\frac{dx_2}{dt} = \frac{(\delta r^{\alpha p-\beta} x_1^{p-1} + b r^{\alpha q} x_1^{q-1} x_2)(r^2 + \alpha x_1^2) - \beta r^{\beta-\alpha} x_2^2}{r^2 + \alpha x_1^2 + \beta x_2^2} x_1; \tag{46.2}$$

and

$$\frac{d\theta}{d\tau} = (1 + \alpha)\left(\delta r^{(\alpha+1)(p+1)-2(\beta+1)}\cos^{p+1}\theta + br^{(\alpha+1)(q+1)-(\beta+1)}\cos^{q+1}\theta\sin\theta\right) - (1 + \beta)\sin^2\theta, \tag{47.1}$$

$$\frac{d\ln r}{d\tau} = \left(1 + \delta r^{(\alpha+1)(p+1)-2(\beta+1)}\cos^{p-1}\theta + br^{(\alpha+1)(q+1)-(\beta+1)}\cos^{q-1}\theta\sin\theta\right)\cos\theta\sin\theta. \tag{47.2}$$

It is obvious that just when $p = 2q + 1$ the complete exponent balance of terms with $\delta$ and $b$ as their coefficients is available; otherwise, only one of them balances with the other term, we will find the zeros by letting $r = 0$ and determine the entry way of streamlines in a parabolic sector which determined by two CDs ($\theta_{p1}$ and $\theta_{p2} > \theta_{p1}$) approaching to the isotropic point. Zhang (1992) [53] further pointed out that in a small neighborhood of the origin any streamline of (43) must tend to the origin spirally or along a fixed direction. The streamline is tangent to $\theta_{p1}$ or $\theta_{p2}$, depending on whether there is a direction $\theta_* \in (\theta_{p1}, \theta_{p2})$ satisfying

$$\frac{d\theta}{d\tau}\big|_{(r,\theta)=(0,\theta_*)} \cdot \frac{d\ln r}{d\tau}\big|_{(r,\theta)=(0,\theta_{p1})} > 0 \text{ or } \frac{d\theta}{d\tau}\big|_{(r,\theta)=(0,\theta_*)} \cdot \frac{d\ln r}{d\tau}\big|_{(r,\theta)=(0,\theta_{p2})} < 0, \tag{48}$$

which indicates that both the circumferential streamline $r = 0, \theta \in (\theta_{p1}, \theta_{p2})$ and the streamline $\theta = \theta_{p1}$ flow in/out the point $(r, \theta) = (0, \theta_{p1})$, or one of the circumferential streamline $r = 0, \theta \in (\theta_{p1}, \theta_{p2})$ and the streamline $\theta = \theta_{p2}$ flows in/out the point $(0, \theta_{p2})$. We will analyze the streamline pattern around the origin as follows:

Case 1: $b = 0$ or $p < 2q + 1$, then we have $(\alpha + 1)(q + 1) - (\beta + 1) > 0$ when
$$(p + 1)(\alpha + 1) = 2(\beta + 1) \leftrightarrow \alpha = 1, \beta = p \tag{49}$$
and the equation (47) can be reduced to
$$\frac{d\theta}{d\tau} = 2\delta\cos^{p+1}\theta - (p+1)\sin^2\theta, \frac{d\ln r}{d\tau} = (1 + \delta\cos^{p-1}\theta)\cos\theta\sin\theta. \tag{50}$$

Case 1.1: when $p$ is odd and $\delta = -1$, there is no separatrix and Ind = 1 from (10) since $e = h = 0$, indicating that the isotropic point is a center with streamlines $x_2^2 + \frac{1}{p+1}x_1^{p+1} = 0$ if $b = 0$, or a focus since $\frac{\partial v_1}{\partial x_1} + \frac{\partial v_2}{\partial x_2} = x_1^q \neq 0$ if $b \neq 0$.

Case 1.2: when $p$ is odd and $\delta = 1$, there are four separatrices for from $\cos^{\frac{p+1}{2}}\theta = \sqrt{\frac{p+1}{2}}\sin\theta$ defined by $\theta = \theta_k \in \left(0, \frac{\pi}{2}\right), \pi - \theta_k, \pi + \theta_k, 2\pi - \theta_k$, in which fluid flowing out, in, out and in, respectively. All four sectors are hyperbolic since the streamlines are externally tangent to the circle centered at the origin. From (10), Ind = $-1$, and the isotropic point is a saddle.

Case 1.3: when $p$ is even ($\delta = 1$), there are two non-collinear separatrices: $\theta = \theta_k \in \left(0, \frac{\pi}{2}\right), 2\pi - \theta_k$ where streamlines flowing out and in, respectively. According to the tangency and flow directions, the streamlines between the separatrices and around the origin can be illustrated as Fig. 4(a) (right). This is a typical cusp with two hyperbolic sectors and so Ind = 0.

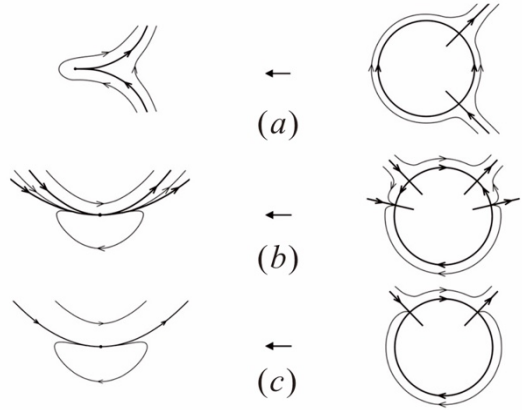

Figure 4. Blow-up of nilpotent isotropic points: (a) a cusp, (b) a four-separatrix elliptical-saddle, (c) a two-separatrix elliptical-saddle.

Case 2: when $b \neq 0$, $p > 2q + 1$, we have $(\alpha + 1)(p + 1) - 2(\beta + 1)(\alpha + 1) > 0$ from $(\alpha + 1)(q + 1) - (\beta + 1) = 0$, and the parameters $\alpha = 0, \beta = q$ reduce the equation (47) to be
$$\frac{d\theta}{d\tau} = \delta r^{p-2q-1}\cos^{p+1}\theta + \cos^{q+1}\theta\sin\theta - (1+q)\sin^2\theta, \frac{d\ln r}{d\tau} = (1 + \delta r^{p-2q-1}\cos^{p-1}\theta + \cos^{q-1}\theta\sin\theta)\cos\theta\sin\theta. \tag{51}$$
Here we encounter a special situation that the lower terms cannot make the origin an isolated singularity. By letting $r = 0$, we obtain $\sin\theta = \frac{1}{q+1}\cos^{q+1}\theta$ and $\sin\theta = 0$ from $\frac{d\theta}{d\tau} = 0$. The former really defines two separatrices $\theta_{1,2}$, but the latter also results in $\frac{d\ln r}{d\tau} = 0$, which comes from the cofactor $x_2$ that is not allowed. Starting from the point $r = 0, \theta = 0$ or $\pi$, $\theta_{3,4}^h$ involving higher order term, called the higher-order separatrix, must be introduced, the solutions $\theta^h(r)$ of $\delta r^{p-2q-1}\cos^{p+1}\theta^h + \cos^{q+1}\theta^h\sin\theta^h - (1+q)\sin^2\theta^h = 0$ with $\theta^h(r = 0) = 0$ or $\pi$ yield two special separatrices.

Case 2.1: when $q$ is odd, we have two separatrices $\theta = \theta_{1,2}$ lying in the quadrants I/II and two higher-order separatrices starting from $(r,\theta) = (0,0)$ and $(0,\pi)$; there are tangency of streamlines to any circle centered at the origin at $\theta = 0, \frac{\pi}{2}, \pi, \frac{3\pi}{2}$, and the radial coordinate of streamlines increases in quadrants I/III and decreases quadrants II/IV. These features mean that the isotropic point could be (1) saddle-node with $\text{Ind} = 0$ if $p = \text{even}$ and the streamlines in parabolic sectors are tangent to $\theta = \pi$ from (48), (2) elliptic-saddle with $\text{Ind} = 1$ if $p = \text{odd}, \delta = -1$ and all of the streamlines in one of the parabolic sectors are tangent to $\theta = 0^+$ and the others are tangent to $\theta = \pi^-$, (3) saddle with $\text{Ind} = -1$ if $p = \text{odd}, \delta = 1$.

Case 2.2: when $q$ is even, there are two separatrices $\theta = \theta_{1,2}$ lying in the quadrants I/III. The isotropic point could be (1) saddle-node with $\text{Ind} = 0$ if $p = \text{even}$, and all of streamlines in parabolic sectors are tangent to $\theta = \pi$, (2) node with $\text{Ind} = 1$ if $p = \text{odd}, \delta = -1$, and all of the streamlines in two parabolic sectors are tangent to $\theta = 0$ and the others are tangent to $\theta = \pi$, (3) saddle with $\text{Ind} = -1$ if $p = \text{odd}, \delta = 1$.

Case 3: when $b \neq 0$, $p = 2q + 1$, taking
$$(p+1)(\alpha+1) = 2(\beta+1) \leftrightarrow \alpha = 0, \beta = \frac{p-1}{2} = q, \tag{52}$$
we have
$$\frac{d\theta}{d\tau} = \delta\cos^{2q+2}\theta + b\cos^{q+1}\theta\sin\theta - (q+1)\sin^2\theta, \tag{53.1}$$
$$\frac{d\ln r}{d\tau} = (1 + \delta\cos^{2q}\theta + b\cos^{q-1}\theta\sin\theta)\cos\theta\sin\theta. \tag{53.2}$$
The structure of the quadratic form (53.1) of $\cos^{q+1}\theta$ and $\sin\theta$ can be clarified by the discriminant
$$\Delta = b^2 + 4\delta(q+1). \tag{54}$$

Case 3.1: if $\Delta < 0$, which is possible only when $\delta = -1$, there is no separatrix, the origin is a focus since $\frac{\partial v_1}{\partial x_1} + \frac{\partial v_2}{\partial x_2} = bx_1^q \neq 0$, and $\text{Ind} = 1$.

Case 3.2: if $\Delta > 0$, and $q$ is odd, $\delta = 1$, we obtain four separatrices $\theta_{1-4}$ in four quadrants from $\cos^{q+1}\theta = \mu_k \sin\theta, k = 1,2$, $\mu_1\mu_2 < 0$, and can determine the origin being a saddle with $\text{Ind} = -1$ from the external tangency and flow directions.

Case 3.3: if $\Delta > 0$, and $q$ is odd, $\delta = -1$, we have four non-collinear separatrices $\theta_{1-4}$ ($0 < \theta_1 < \theta_2 < \frac{\pi}{2} < \theta_3 < \theta_4 < \pi$) from $\cos^{q+1}\theta = \mu_i \sin\theta$, $\mu_1, \mu_2 > 0$ since $b > 0$. As shown in Fig. 4(b), one hyperbolic sector and an elliptic sector separated by two parabolic sectors yields an elliptic-saddle singularity with $\text{Ind} = 1$, and according to the criterion presented in (48), all of the streamlines in one of parabolic sectors are tangent to $\theta = \theta_1^+$ and the others are tangent to $\theta = \theta_4^-$.

Case 3.4: if $\Delta > 0$, and $q$ is even, $\delta = 1$, we obtain a saddle with four separatrices in four quadrants, and $\text{Ind} = -1$.

Case 3.5: for $\Delta > 0$, $q$ is even, $\delta = -1$, we have four separatrices $\theta_{1-4}$ ($0 < \theta_1 < \theta_2 < \frac{\pi}{2}$, $\pi < \theta_3 < \theta_4 < \frac{3}{2}\pi$), and get a node with $\text{Ind} = 1$ from four parabolic sectors, and all of the streamlines in two parabolic sectors are tangent to $\theta = \theta_1$ and the others are tangent to $\theta = \theta_3$.

Case 3.6: if $\Delta = 0$ ($\delta = -1$), and $q$ is odd, from $\cos^{q+1}\theta - \frac{b}{2}\sin\theta = 0$, we have two separatrices in the quadrants I/II, and derive a hyperbolic sector plus an elliptic sector, and so obtain an elliptic-saddle singularity with $\text{Ind} = 1$, as shown in Fig. 4(c).

Case 3.7: if $\Delta = 0$ ($\delta = -1$), and $q$ is even, also from $\cos^{q+1}\theta - \frac{b}{2}\sin\theta = 0$, we have two separatrices $0 < \theta_1 < \frac{\pi}{2}, \pi < \theta_2 < \frac{3}{2}\pi$, derive two parabolic sectors resulting in a node with $\text{Ind} = 1$, and all of the streamlines in one of the parabolic sectors are tangent to $\theta = \theta_1^-$ and the others are tangent to $\theta = \theta_2^-$ in another parabolic sector.

In summary, we carry out the classification of semi-elementary and nilpotent velocity fields, which can be listed in Table 3 and Table 4, respectively. It should be pointed out that the nilpotent cases previously analyzed have the same type of LSP, such as saddle, but they cannot be merged through spatiotemporal transformations,

Table 4 shows the merged results in order to make the classification concise. By comparing with the LSP of elemental singularities, in semi-elementary velocity fields a new type, saddle-node, has emerged while three new types, namely cusp (two separatrices), saddle-node (four separatrices) and elliptic-saddle (four/two separatrices), have emerged in nilpotent velocity fields.

Table 3. LSP Types of semi-elementary velocity fields

| Equation | Parameters | Index | Sectors/Separatrices | Type | LSP |
|---|---|---|---|---|---|
| $\begin{cases}\dot{x}_1 = x_1 \\ \dot{x}_2 = \delta x_2^m\end{cases}$ | $m = 2k+1, \delta = 1$ | 1 | 4/4 | node | Fig. 3(a) |
|  | $m = 2k+1, \delta = -1$ | −1 | 4/4 | saddle | Fig. 3(b) |
|  | $m = 2k, \delta = 1$ | 0 | 4/4 | saddle-node | Fig. 3(c) |

Table 4. LSP Types of nilpotent velocity fields

| Equations | Index | Sectors/Separatrices | Parameters | Type | Pattern |
|---|---|---|---|---|---|
| $\begin{cases}\dot{x}_1 = x_2 \\ \dot{x}_2 = \delta x_1^p + b x_1^q x_2\end{cases}$ | 0 | 2/2 | $2k = p < 2q + 1$ | cusp | 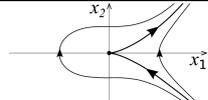 |
|  | 0 | 4/4 | $2k = p > 2q + 1$ | saddle-node | 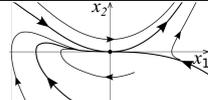 |
|  | -1 | 4/4 | $2k + 1 = p, \delta = 1$ | saddle | 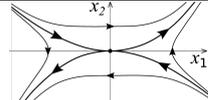 |
|  | 1 | 4/4 | $2k + 1 = p \geq 2q + 1, \delta = -1$, $q = 2l + 1$ (for $p = 2q + 1, \Delta > 0$) | elliptic-saddle | 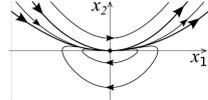 |
|  | 1 | 4/4 | $2k + 1 = p \geq 2q + 1, \delta = -1$, $q = 2l$ (for $p = 2q + 1, \Delta > 0$) | node | 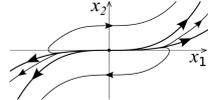 |
|  | 1 | 0/0 | $2k + 1 = p < 2q + 1, \delta = -1$, $b = 0$ | center | 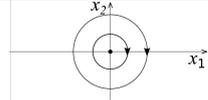 |
|  | 1 | 0/0 | $2k + 1 = p \leq 2q + 1, \delta = -1$, $b \neq 0$ (for $p = 2q + 1, \Delta < 0$) | focus | 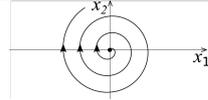 |
|  | 1 | 2/2 | $p = 2q + 1, \delta = -1, \Delta = 0, q = 2l$ | node | 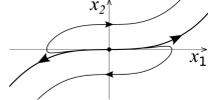 |
|  | 1 | 2/2 | $p = 2q + 1, \delta = -1, \Delta = 0$, $q = 2l + 1$ | elliptic-saddle | 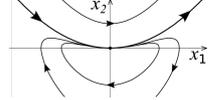 |

## 5. Index of nonlinear velocity field with nonzero linear part

In the previous section, from $\frac{d\theta}{dt}$ or that after blow-up, we determine the separatrices in the neighbourhood of the isotropic point, mostly depending on the circumferential coordinate. When the lowest terms of velocity have cofactor, the higher order separatrices involving the radial coordinate must be considered. The separatrices divide the neighbourhood of the isotropic point into several sectors. Then, it is available to determine the flow directions the separatrices from the sign of $\frac{d\ln r}{dt}$, and judge from the possible zero points of $\frac{d\ln r}{dt}$ whether the streamlines in different sectors are tangent to a circle centered at the origin or not. When the tangent points exist, we can further determine whether it is inscribed or circumscribed by determining whether the radial coordinates of the streamline on both sides of the tangent point increase or decrease. For a parabolic sector where two separatrices are streamlines that flow out or in in the same time, we can make clear which one is tangent to all streamlines in the sector. When all these qualitative characteristics are clear, we can decide the types of sectors, and draw up the streamline pattern. And the index of the isotropic point is easy to calculate from the Bendixson formula (9).

However, there exists a more direct method for calculating the index of an isotropic point. Using the complex representation $v = v_1 + \iota v_2$ of the velocity, we have from the definition (4)

$$\text{Ind}(v, \mathcal{L}) = \frac{1}{2\pi} \oint_{\mathcal{L}} \frac{\text{Im}(\bar{v}dv)}{V^2} = \frac{1}{2\pi} \text{Im} \oint_{\mathcal{L}} \frac{\bar{v}dv}{\bar{v}v} = \frac{1}{2\pi} \text{Im} \oint_{\mathcal{L}} d\ln v. \tag{55}$$

when the isotropic point is the unique singularity in $\mathcal{L}$ and there are no singular points on $\mathcal{L}$. No loss of generality, we take $\mathcal{L}$ to be a circle centered at the isotropic point and with a radius that can guarantee the uniqueness of singularity in and on the circle. Assume that the velocity is analytic in the plane, we have the polynomial expansion (3), and it is proved that if the isotropic point of the velocity is also that of $v = P_m + \iota Q_n$ ($m \leq n$), their indices are the same (Zhang, 1992) [53]. From the context of this paper, we have $m \equiv 1$. When $Q_n$ have $P_1$ as its factor, higher term needs to be considered.

For the linear velocity fields with non-degenerate coefficient determinant, making use of

$$v = \bar{d}^{(0)}z + d^{(2)}\bar{z}, \qquad d^{(0)} = \vartheta + \iota\omega, d^{(2)} = \tau e^{2\iota\alpha}, |d^{(0)}| + |d^{(2)}| > 0, \tag{56}$$

we have the coefficients (12)$_2$ as

$$d_{11} = \vartheta + \tau\cos2\alpha, d_{12} = \omega + \tau\sin2\alpha, d_{21} = -\omega + \tau\sin2\alpha, d_{22} = \vartheta - \tau\cos2\alpha. \tag{57}$$

Substituting (56) into (55), and let $\mathcal{L}$ be a circle with radius $r_0$ small enough such that $z = r_0\eta = r_0 e^{\iota\theta}$, yields

$$\text{Ind}(v, \mathcal{L}) = \frac{1}{2\pi} \text{Im} \oint_{|\eta|=1} d\ln(\bar{d}^{(0)}\eta + d^{(2)}\eta^{-1}) = \text{sgn}(|d^{(0)}| - |d^{(2)}|) = \text{sgn}(J_2), \tag{58}$$

since $J_2 = \vartheta^2 - \tau^2 + \omega^2 = d^{(0)}\bar{d}^{(0)} - d^{(2)}\bar{d}^{(2)}$, while $J_2 = 0$ means two velocity components are linear dependent. This formula can reach directly from the properties (5)$_1$ and (6) by taking $v_1 = x_1, v_2 = x_2$.

As pointed out, when $J_2 = 0$, higher order terms should be considered, we investigate the indices of the semi-elementary velocity (39) and the nilpotent velocity (43) in the following. Since the origin is the unique singular point of two velocity fields in the plane, the close curve $\mathcal{L}$ can be chosen to be a unit circle. For the semi-elementary case, we have

$$\text{Ind}_O = \text{Ind}(v, \mathcal{L}) = \frac{1}{2\pi} \text{Im} \oint_{\mathcal{L}} d\ln v = \frac{\text{sgn}(\delta)}{2\pi} \text{Im} \int_0^{2\pi} d\ln(\cos\theta + \iota\sin^m\theta)$$

$$= \frac{\text{sgn}(\delta)}{2\pi} \text{Im} \int_0^{\pi} d\ln(\cos\theta + \iota\sin^m\theta) + d\ln(\cos\theta - (-1)^m \iota\sin^m\theta)$$

$$= \begin{cases} \frac{\text{sgn}(\delta)}{\pi} \text{Im} \int_0^{\pi} d\ln(\cos\theta + \iota\sin^{2k+1}\theta) & ,m = 2k+1, \\ 0 & ,m = 2k. \end{cases} \tag{59}$$

where use is made of the property (5)$_3$. Write $\cos\theta + \iota\sin^{2k+1}\theta$ in modulus and argument representation $B(\theta)e^{\iota\phi(\theta)}$, we find the range of $\phi(\theta)$ is the same as that of $\theta$, say between $(0, \pi)$, and derive the following result:

$$\text{Ind}_O = \begin{cases} \text{sgn}(\delta) & ,m = 2k+1, \\ 0 & ,m = 2k. \end{cases} \tag{60}$$

For the nilpotent velocity, first when $p < q+1$ or $b = 0$, the principal equation becomes

$$\dot{x}_1 = x_2, \dot{x}_2 = \delta x_1^p, \tag{61}$$

and similar consideration resulting in

$$\text{Ind}_O = \frac{\text{sgn}(\delta)}{2\pi} \text{Im} \int_0^{2\pi} d\ln(\sin\theta + \iota\cos^p\theta) = \frac{\text{sgn}(\delta)}{2\pi} \text{Im} \int_0^{2\pi} d\ln(\iota e^{-\iota\varphi(\theta)}) = \begin{cases} 0 & ,p = 2k, \\ -\text{sgn}(\delta) & ,p = 2k+1. \end{cases} \tag{62}$$

In general, by setting $b' = b/\delta$, we can write the formula of nilpotent velocity as

$$\text{Ind}_O = \frac{1}{2\pi} \text{Im} \oint_{\mathcal{L}} d\ln v = \frac{\text{sgn}(\delta)}{2\pi} \text{Im} \int_0^{2\pi} d\ln[\sin\theta + \iota Y(\theta)], Y(\theta) = \cos^p\theta + b'\sin\theta\cos^q\theta, \tag{63}$$

From the previous examples, we know that $\text{Ind}_O = 0$ if $Y(\theta + \pi) = Y(\theta)$, which becomes true when $p = $ even, $q = $ odd. For the cases not like this, following the idea of Gao (1962) [66], the results of (63) are simply determined by the signs of $Y(\theta)$ at the zero points of $\sin\theta$, namely $\theta = 0, \pi$: (1) $\text{Ind}_O = 0$ if the two signs are the same to each other; (2) $\text{Ind}_O = \mp\text{sgn}(\delta)$ if the signs are the same/opposite as those of $\cos\theta$ at the two points. According to this logic, we can omit the factor that keeps positive at $\theta = 0, \pi$, and have the following derivations.

(1) when $p = q + 1$,

$$\text{Ind}_O = \frac{1}{2\pi}\text{Im}\int_0^{2\pi} d\ln[\sin\theta + \iota(\delta\cos\theta + b\sin\theta)\cos^q\theta] = \begin{cases} 0 & , p = 2k, \\ \frac{1}{2\pi}\text{Im}\int_0^{2\pi} d\ln(\sin\theta + \iota\delta\cos\theta) = -\text{sgn}(\delta) & , p = 2k + 1; \end{cases} \quad (64)$$

(2) when $p > q + 1$, $q = $ even,

$$\text{Ind}_O = \frac{1}{\pi}\text{Im}\int_{-\frac{\pi}{2}}^{\frac{\pi}{2}} d\ln[\sin\theta + \iota(\delta\cos^{p-q}\theta + b\sin\theta)] = \begin{cases} 0 & , p = 2k, \\ -\text{sgn}(\delta) & , p = 2k + 1; \end{cases} \quad (65)$$

(3) when $p > q + 1$, $q = $ odd, $p = $ odd

$$\text{Ind}_O = \frac{1}{2\pi}\text{Im}\int_{-\pi}^{\pi} d\ln[\sin\theta + \iota(\delta\cos^{p-q}\theta + b\sin\theta)\cos^q\theta] = \frac{\text{sgn}(\delta)}{2\pi}\text{Im}\int_{-\pi}^{\pi} d\ln(\sin\theta + \iota\cos^q\theta) = -\text{sgn}(\delta). \quad (66)$$

In summary, we conclude the results for the nilpotent velocity field as

$$\text{Ind}_O = \begin{cases} 0 & , p = \text{even}, \\ -\text{sgn}(\delta) & , p = \text{odd}. \end{cases} \quad (67)$$

The results (60) and (67) of the index for nonlinear velocity fields with nonzero linear part coincide with those listed in Table 3 and Table 4, respectively.

## 6. Discussions and Conclusions

It is well known that the complexity of steady flows comes from the streamline pattern, which reflects the contact state of fluid during its transport process, and doesn't obey the invariance under the Galilean transformation. The fundamental of understanding the streamline pattern is the classification of the LSP, which constitutes the research scheme of this paper. In flows or their simulations, since the velocity field is usually nonlinear, it is not enough to investigate the streamline pattern of linear velocity fields or from the velocity gradient. A typical higher order isotropic point appears in the creeping flow past two cylinders in tandem, as figure no. 15 in the famous book of van Dyke [67].

The streamline pattern is connected with velocity direction, while the speed is dynamic but has no contribution to the streamline pattern. Topologically, vortex, as a special type of LSP called center or spiral, must be combined with other types of LSPs to form a globally complex flow structure. Dynamically, isotropic points are easy to happen in fluid flows, but the mathematical study of streamline patterns of nonlinear velocity fields is not easy. Up to now, there are relatively complete results on the LSPs of planar nonlinear velocity fields with nonzero linear part and the global streamline pattern of planar quadratic velocity fields.

In this paper, we propose a novel formulation of qualitative equivalence, namely the invariance under spatiotemporal transformations, to build up the quasi-real Schur form for classification of linear velocity fields, and starting from this analyze the normal forms of nonlinear velocity fields with nonzero linear part, where for the nilpotent case use is made of the polar blow-up technique; and finally we provide a new method to calculate the index of polynomial velocity fields at its isotropic point. Most conclusions of this article are not new and are scattered to some extent in various literatures. We adopt the name LSP instead of local phase portrait to sort out and report these results, which will help fluid mechanics community gain a comprehensive understanding of the complexity of this problem, and a promote in-depth research in this field will be expected.

**Disclosure statement**

No potential conflict of interest was reported by the authors.

**Funding**

This work is supported by the Open Fund of Jiangxi's Key Laboratory of Aircraft Design and Aerodynamic Simulation, grant no. E202280266.

**Acknowledgments:**

Jian Gao acknowledges Prof. Wennan Zou for his supervision in this study.

**Data Availability Statement**

The data that support the findings of this study are available from the author upon reasonable request.

# References

1. U. Frisch, S.A. Orszag. Turbulence: challenges for theory and experiment. *Phys. Today*, 43(1): 24-32 (1990).
2. G.I. Barenblantt, A.J. Chorin. Turbulence: an old challenge and new perspectives. *Meccanica*, 33: 445-468 (1998).
3. G.K. Batchelor. *An Introduction to Fluid Dynamics*. Cambridge University Press (1970).
4. V.I. Arnold, B.A. Khesin. *Topological Method in Hydrodynamics*. 2nd ed., Springer (2001).
5. H.K. Moffat. Some topological aspects of fluid dynamics. *J. Fluid Mech.*, 914, 1-53 (2021).
6. A.I. Murdoch. *Physical Foundations of Continuum Mechanics*. Cambridge Univ. Pr. (2012).
7. W.N. Zou. Reconstructing fluid dynamics with micro-finite element. arXiv:1708.06059v1 (2017).
8. D. Bernoulli. *Hydrodynamics*. 1738, (English edition, Dover Publications, Inc. New York, 1968).
9. J.S. Calero. *The Genesis of Fluid Mechanics 1640-1780*. Springer (2008).
10. L. Euler. German translation and commentary on the New principles of Gunnery by B. Robins, Berlin, 1745.
11. C. Truesdell. *Essays in the History of Mechanics*. Springer-Verlag, Berlin (1968).
12. W.J.M. Rankine. 1871. On the mathematical theory of streamlines, especially those with four foci and upwards. PT, 161: 267-303, https://doi.org/10.1098/rstl.1871.0011 (1871).
13. J.M. Délery. Robert Legendre and Henri Werlé: Toward the Elucidation of three-dimensional separation. *Annu. Rev. Fluid Mech.*, 33: 129-54 (2001).
14. M. Tobak, D.J. Peake. Topology of three-dimensional separated flows. *Annu. Rev. Fluid Mech.*, 14: 61-85 (1982).
15. A.E. Perry, M.S. Chong. A description of eddying motions and flow patterns using critical-point concepts. *Annu. Rev. Fluid Mech.*, 19(1): 125-155 (1987).
16. P.G. Bakker. Bifurcations in Flow Patterns. Vol. 2 of *Nonlinear Topics in the Mathematical Sciences*, Edited by M.S. Berger, Springer, (1991).
17. J. Guckenheimer, P. Holmes. *Nonlinear Oscillations, Dynamical Systems and Bifurcations of Vector Fields*. Springer, (1983).
18. M. Brøns. Topological fluid dynamics of interfacial flows. *Physics of Fluids*, 6, 2730 (1994).
19. M. Brøns. Streamline topology: patterns in fluid flows and their bifurcations. *Advances in Applied Mechanics*, 41: 1-42 (2007).
20. M. Brøns and Hartnack, Streamline topologies near simple degenerate critical points in two-dimensional flow away from boundaries. *Phys. Fluids*, 11(2): 314-324 (1999).
21. M. Brøns, L.K. Voigt, J.N. Sørensen. Streamline topology of steady axisymmetric vortex breakdown in a cylinder with co- and counter-rotating end-covers. *J. Fluid Mech.*, 401: 275-292 (1999).
22. M. Brøns, L.K. Voigt, J.N. Sørensen. Topology of vortex breakdown bubbles in a cylinder with rotating bottom and free surface. *J. Fluid Mech.*, 428: 133-148 (2001).
23. J.N. Hartnack. Streamline topologies near a fixed wall using normal forms. *Acta Mech.*, 136(1-2): 55-75 (1999a).
24. J.N. Hartnack. *Structural Changes in Incompressible Flow Patterns*. PhD thesis, Department of Mathematics, Technical University of Denmark, 1999b.
25. J. Jiménez-Lozano, M. Sen. Streamline topologies of two-dimensional peristaltic flow and their bifurcations. *Chemical Engineering and Processing*, 49: 704-715 (2010).
26. A. Deliceoglu. Topology of two-dimensional flow associated with degenerate dividing streamline on a free surface. *Euro. J. Appl. Math.*, 24: 77-101 (2013).
27. T. Yokoyama, T. Sakajo. Word representation of streamline topologies for structurally stable vortex flows in multiply connected domains. *Proc R Soc A*, 469: 20120558 (2015).

28. T. Sakajo, T. Yokoyama. Tree representations of streamline topologies of structurally stable 2D incompressible flows. *IMA J. Appl. Math.*, 00: 1-32 (2018).
29. K. Bajer, H.K. Moffatt. On a class of steady confined Stokes flows with chaotic streamlines. *J. Fluid Mech.*, 212: 337-363 (1990).
30. M. Brøns. Streamline pattern and their bifurcations using methods from dynamical systems, in *An Introduction to the Geometry and Topology of Fluid Flows*, edited by RL Ricca, 167-182 (2001).
31. G.C. Layek. *An Introduction to Dynamical Systems and Chaos*. Springer (2015).
32. H. Poincaré. Mémoire sur les courbes définies par une équation différentielle. *J. de Math.*, 2: 151-217 (1886).
33. G.D. Birkhoff. Quelques théorèmes sur le mouvement des systèmes dynamiques. *Bull. Soc. Math.* France 40: 305-323 (1912).
34. G.D. Birkhoff. *Dynamical Systems*. Amer. Math. Soc., Providence, RI. (1927).
35. A.A. Andronov, A.A. Vitt, S.E. Khaikin. *Theory of Oscillators*. Dover, New York, 1966.
36. M. Brin, G. Stuck. *Introduction to Dynamical Systems*. Cambridge (2001).
37. N.G. Lloyd. Limit cycles of polynomial systems - some recent development. In *New Directions in Dynamical Systems*, London Math. Soc. Lect. Notes Ser. 127: 192-234 (1988).
38. S. Wiggins. *Introduction to Applied Nonlinear Dynamical Systems and Chaos*. Second Edition, Springer, 1990.
39. J. Llibre, A.E. Teruel. *Introduction to the Qualitative Theory of Differential Systems: Planar, Symmetric and Continuous Piecewise Linear Systems*. Birkhäuser Advanced Texts, Springer, 2014.
40. V.I. Arnold. *Ordinary Differential Equations*. 3rd Ed., Translated from the Russian by R. Cooke, Springer-Verlag, Page 195, 1992.
41. M.S. Chong, A.E. Perry, B.J. Cantwell. A general classification of three-dimensional flow fields. *Phys. Fluids A*, 2, 765-777 (1990).
42. F. Verhulst. *Nonlinear Differential Equations and Dynamical Systems*. Springer-Verlag Berlin Heidelberg, 1996.
43. M.W. Hirsch, S. Smale, R.L. Devany. *Differential Equations, Dynamical Systems, and An Introduction to Chaos*. 2nd Ed., Elsevier Academic Press, 2004.
44. W.N. Zou, X.Y. Xu, C.X. Tang. Spiral streamline pattern around a critical point: Its dual directivity and effective characterization by right eigen representation. *Phys. Fluids*, 33, 067102 (2021).
45. E.N. Lorenz. Deterministic non-periodic flow. *J. Atmos. Sci.*, 20: 130-141 (1963).
46. D. Ruelle, F. Takens. On the nature of turbulence. *Commun. Math. Phys.*, 20: 167-192 (1971).
47. W.G. Kelley, A.C. Peterson. *The Theory of Differential Equations: Classical and Qualitative*. 2nd Ed., Springer-Verlag New York, 2010.
48. Y.Q. Ye. *Theory of Limit Cycles*. Translations of Mathematical Monographs, Vol. 66, American Mathematical Society, (1986).
49. J.B. Li. Hilbert's 16th problem and bifurcations of planar polynomial vector fields. *Int. J. Bif. Chaos*, 13(1): 47-106 (2003).
50. F. Dumortier, J. Llibre, J.C. Artés. *Qualitative Theory of Planar Differential Systems*. Springer, 2006.
51. R. Legendre. Séparation de l' écoulement laminaire tridimensionnel. *La Rech. Aéronaut.*, 54: 3-8 (1956).
52. M. Tabak. D.J. Peake. Topology of three-dimensional separated flows. *Annu. Rev. Fluid Mech.*, 14: 61-85 (1982).
53. Z.F. Zhang, T.R. Ding, W.Z. Huang, Z.X. Dong. *Qualitative Theory of Differential Equations*. Translations of Mathematical Monographs, Vol. 101, American Mathematical Society, 1992.
54. J.C. Artés, J. Llibre, D. Schlomiuk, N. Vulpe. *Geometric Configurations of Singularities of Planar Polynomial Differential Systems: A Global Classification in the Quadratic Case*. Springer, (2021).
55. R. Grimshaw. *Nonlinear Ordinary Differential Equations*. Blackwell Scientific Pub., 1990.
56. V.V. Nemytskii, V.V. Stepanov. *Qualitative Theory of Differential Equations*. Princeton Univ. Press (1960) (Translated from Russian).
57. J. Cronin. *Ordinary Differential Equations: Introduction and Qualitative Theory*. Third Edition, CRC Press, 2007.
58. Q.B. Jiang, J. Llibre. Qualitative classification of singular points. *Qualitative Theory of Dynamical Systems*, 6: 87-167 (2005).
59. J.C. Artés, J. Llibre, D. Schlomiuk, N. Vulpe. From topological to geometric equivalence in the classification of singularities at infinity for quadratic vector fields. *Rocky Mountain J. of Math.*, 45(1): 29-113 (2015).
60. V.I. Arnold. *Geometrical Methods in the Theory of Ordinary Differential Equations*. Second Edition, Springer, 1988.
61. P.B. Kahn, Y. Zarmi. *Nonlinear Dynamics Exploration Through Normal Forms*. Dover, 1998.
62. A.H. Nayfeh. *The Method of Normal Forms*. Second, updated and enlarged edition, Wiley-VCH, 2011.
63. J. Zhou, R. Adrian, S. Balachandar, K.M. Kendall. Mechanisms for generating coherent packets of hairpin vortices in channel flow. *J. Fluid Mech.*, 387: 353-396 (1999).

64. Z. Li, X.W. Zhang, F. He. Evaluation of vortex criteria by virtue of the quadruple decomposition of velocity gradient tensor (in Chinese). *Acta Phys. Sin.*, 63(5), 054704 (2014).
65. C.Q. Liu, Y.S. Gao, S.L. Tian, X.R. Dong. Rortex - A new vortex vector definition and vorticity tensor and vector decompositions. *Phys. Fluids*, 30, 035103 (2018).
66. W.X. Gao. Indices for planar critical points (in Chinese). *Symposium of Differential Equations*, Department of Mathematics and Mechanics, Beijing University, 189-198 (1962).
67. M. van Dyke. *An Album of Fluid Motion*. The Parabolic Press, 1982.